\documentclass[a4paper,english,prl]{revtex4}
\usepackage[T1]{fontenc}
\usepackage[latin1]{inputenc}
\usepackage{amsmath}
\usepackage{graphicx}
\usepackage{amssymb}

\makeatletter

\providecommand{\tabularnewline}{\\}

\usepackage{graphicx}
\usepackage{dcolumn}
\usepackage{bm}
\usepackage{mathrsfs}

\usepackage{babel}
\makeatother
\begin{document}

\title{Magnetoresistance in metals with embedded magnetic nanoclusters.}

\author{O. Tsyplyatyev}

\affiliation{Physics Department, Lancaster University, Lancaster LA1 4YB, UK}

\author{Vladimir I. Fal'ko}

\affiliation{Physics Department, Lancaster University, Lancaster LA1 4YB, UK}

\begin{abstract}
We propose a kinetic theory of transport in metals embedded with ferromagnetic
nanoclusters, describing the dependence of the form of magnetoresistance
on anisotropy characteristics of the ensemble of clusters and the
influence of the electron spin depolarisation by clusters. We note
that this effect is strongest when all the clusters have the same
intrinsic easy-axis anisotropy.
\end{abstract}

\date{\today}

\maketitle
Numerous applications of magnetoresistance (MR) phenomenon fuel a
continuous search for new materials, including various hybrid systems
\cite{key-26}. In this publication, we propose a kinetic theory of
high temperature MR in a normal metal embedded with ferromagnetic
metallic nano-clusters (FmnC) \cite{key-11}. 

The presented below analysis extends the earlier phenomenological
study of MR effect due to FmnC \cite{key-30} developed in application
to granular normal-ferromagnetic metal films \cite{key-29}. Following
Zhang \cite{key-30}, we study clusters using classical spin approximation.
We also assume that both the magnetisation and the electron exchange
spin-splitting $J$ are homogeneous within each cluster (due to their
small size, typically $10^{2}-10^{3}$ atoms per cluster \cite{key-11,key-35,key-34})
and we characterise the ensemble of \mbox{FmnC} using the angular
distribution of their magnetic moments $\mathbf{m}=\mu\mathbf{l}$,
which depends on intrinsic anisotropy, external magnetic field and
temperature. This study is restricted to materials with such separation
between clusters that one can neglect correlations between them and
assume that all clusters are of the same radius, $r_{cl}$. 

Due to band mismatch between underlying materials, each FmnC would
generate a scalar potential well and have a different electron mass
from that in the normal metal, in addition to a spin dependent (exchange)
perturbation. An interplay between all above-mentioned factors makes
electronic scattering from each individual FmnC spin-dependent and
the resistance of a composite material sensitive to the mean degree
of polarisation of clusters $\left\langle l_{z}\right\rangle $. In
agreement with earlier publications \cite{key-30,key-29,key-33},
we find that in the low field regime ($\mu B/kT\ll1$) MR is quadratic
in magnetic field, $\Delta\left(B\right)=\frac{R\left(B\right)-R\left(0\right)}{R\left(0\right)}\sim-B^{2}$,
whereas the detailed form of MR is shown below to depend on both mean
polarisation of FmnCs and its variance. We also analyse the influence
of magnetic anisotropy on MR, $\Delta\left(B\right)$ and find that
the effect is strongest when all clusters in the ensemble have collinear
easy-axis anisotropy.

To describe the MR in a metal with FmnCs we study the quantum kinetic
equation,\[
\partial_{t}\hat{\rho}+\mathbf{v}\cdot\nabla\hat{\rho}+e\left(\mathbf{E}\cdot\mathbf{v}\right)\partial_{\epsilon}\hat{\rho}-\frac{\omega_{B}}{2}\left[\hat{\sigma}_{3},\hat{\rho}\right]=I_{0}\left[\hat{\rho}\right]+\left\langle I\left[\hat{\rho}\right]\right\rangle \]
 for the electron spin density matrix $\hat{\rho}=\rho_{0}+\sum_{i}\rho_{i}\hat{\sigma}_{i}$
approximated by $\rho_{\alpha}\left(\mathbf{p}\right)=\breve{\rho}_{\alpha}\left(p\right)+\frac{\mathbf{p}}{p}\cdot\boldsymbol{\eta}_{\alpha}\left(p\right)$,
where $\boldsymbol{\sigma}=\left(\hat{\sigma}_{1},\hat{\sigma}_{2},\hat{\sigma}_{3}\right)$
are Pauli matrices, $\omega_{B}=Be\hbar/mc$ is the electron spin
precession frequency in normal metal and $\breve{\rho}$ is the isotropic
part of the electron distribution in momentum space. All details of
the electron spin and phase evolution inside clusters are incorporated
into the collision integral $I\left[\hat{\rho}\right]$, whereas momentum
relaxation due to non-magnetic impurities and phonons in the normal
metal is taken into account by $I_{0}\left[\hat{\rho}\right]=\left[\hat{\rho}\left(\mathbf{p}\right)-\breve{\rho}\left(p\right)\right]\tau^{-1}$. 

One FmnC produces the following scattering matrix, \begin{equation}
\hat{S}=a+b\left(\boldsymbol{\sigma}\cdot\mathbf{l}\right);\;\,\, a=\frac{S^{+}+S^{-}}{2},b=\frac{S^{-}-S^{+}}{2},\label{eq:f_matrix}\end{equation}
where $S_{\mathbf{pp'}}^{+}$ and $S_{\mathbf{pp'}}^{-}$ characterise
the angular dependence of scattering matrices of electrons with spin
parallel and antiparallel to the cluster polarisation $\mathbf{l}$
{[}$\mathbf{p}$, $\mathbf{p}'$ are electron momenta before and after
scattering{]}. The collision integral taking into account a group
of collinearly polarised clusters with the concentration $n\left(\mathbf{l}\right)$
has the form

\begin{equation}
I_{\mathbf{l}}\left[\hat{\rho}\right]=\frac{n\left(\mathbf{l}\right)\hbar^{2}}{4mp}\int d^{3}p'\bigg[\hat{S}_{\mathbf{pp'}}\hat{\rho}_{\mathbf{p'}}\hat{S}_{\mathbf{p'p}}^{\dagger}-\frac{1}{2}\left(\hat{S}_{\mathbf{pp'}}\hat{S}_{\mathbf{p'p}}^{\dagger}\hat{\rho}_{\mathbf{p}}+\hat{\rho}_{\mathbf{p}}\hat{S}_{\mathbf{pp'}}\hat{S}_{\mathbf{p'p}}^{\dagger}\right)\bigg]\label{eq:kin_eq_f}\end{equation}
where $v$ is Fermi velocity. This expression incorporates the kinetics
of both electron scattering and spin precession, in particular since
$S$ matrix takes into account spin-dependent forward scattering\textit{.}
Due to the angular distribution of cluster polarisations, the ensemble-averaged
collision integral, $\left\langle I\left[\hat{\rho}\right]\right\rangle $
would also describe spin relaxation of electrons. All of the above-mentioned
three generic elements of electron spin kinetics would be present
even if the distribution of cluster magnetic moments is axially symmetric
about the direction of an external magnetic field $\mathbf{B}=B\mathbf{e}_{z}$.
Therefore, in the following analysis we assume that $\left\langle l_{x}\right\rangle =\left\langle l_{y}\right\rangle =0$
and the tensor $\left\langle l_{i}l_{j}\right\rangle $ is diagonal
with $\left\langle l_{x}^{2}\right\rangle =\left\langle l_{y}^{2}\right\rangle =1-\left\langle l_{z}^{2}\right\rangle $.
The resulting ensemble-averaged collision integral reads as\begin{multline*}
\left\langle I\left[\hat{\rho}\right]\right\rangle =\frac{n_{c}\hbar^{2}}{4mp}\int d^{3}p'\Bigg\{\left(ab^{*}+a^{*}b\right)\left(\rho_{3}'-\rho_{3}\right)\left\langle l_{z}\right\rangle \\
+\left(\left|a\right|^{2}+\left|b\right|^{2}\right)\left[\left(\rho_{0}'-\rho_{0}\right)+\sum_{i}\hat{\sigma}_{i}\left(\rho_{i}'-\rho_{i}\right)\right]\\
-2\left|b\right|^{2}\sum_{i}\hat{\sigma}_{i}\rho_{i}'+\imath\left(a^{*}b-ab^{*}\right)\left(\rho_{2}'\left\langle l_{z}\right\rangle \hat{\sigma}_{1}-\rho_{1}\left\langle l_{z}\right\rangle \hat{\sigma}_{2}\right)\\
+\left(ab^{*}+a^{*}b\right)\left(\rho_{0}'-\rho_{0}\right)\left\langle l_{z}\right\rangle \hat{\sigma}_{3}+2\left|b\right|^{2}\sum_{i}\rho_{i}'\left\langle l_{i}^{2}\right\rangle \hat{\sigma}_{i}\Bigg\}\end{multline*}
where density matrix $\hat{\rho}$ was decomposed into a singlet($\rho_{0}$)
and triplet($\rho_{i}$) parts {[}here $\rho=\rho\left(\mathbf{p}\right)$,
$\rho'=\rho\left(\mathbf{p'}\right)]$, which decouples the kinetic
equation into two groups. The first group describes the evolution
of only diagonal density matrix components and involves momentum ($\tau_{p\pm}^{-1}$)
and spin ($\tau_{sn}^{-1}$) relaxation rates,

\begin{alignat}{1}
 & \begin{array}{l}
\partial_{t}\rho_{0}+\frac{v}{3}\nabla\cdot\boldsymbol{\eta}_{0}=0\end{array}\nonumber \\
 & \partial_{t}\rho_{3}+\frac{v}{3}\nabla\cdot\boldsymbol{\eta}_{3}=-\left(1-\left\langle l_{z}^{2}\right\rangle \right)\tau_{s0}^{-1}\rho_{3}\label{eq:rho03_set}\\
 & v\nabla\rho_{0}+ev\mathbf{E}\partial_{\epsilon}\rho_{T}\left(\epsilon\right)=-\tau_{p+}^{-1}\boldsymbol{\eta}_{0}-\tau_{p-}^{-1}\left\langle l_{z}\right\rangle \boldsymbol{\eta}_{3}\nonumber \\
 & v\nabla\rho_{3}=-\tau_{p-}^{-1}\left\langle l_{z}\right\rangle \boldsymbol{\eta}_{0}-\left(\tau_{p+}^{-1}+\tau_{s1}^{-1}\left(1-\left\langle l_{z}^{2}\right\rangle \right)\right)\boldsymbol{\eta}_{3},\nonumber \\
 & \begin{array}{l}
\tau_{p+}^{-1}=\frac{n_{c}\hbar^{2}\pi}{4mp}\int d\theta\sin\theta\left(\left|S^{+}\right|^{2}+\left|S^{-}\right|^{2}\right)\left(1-\cos\theta\right)+\frac{1}{\tau}\\
\tau_{p-}^{-1}=\frac{n_{c}\hbar^{2}\pi}{4mp}\int d\theta\sin\theta\left(\left|S^{+}\right|^{2}-\left|S^{-}\right|^{2}\right)\left(1-\cos\theta\right),\\
\tau_{sn}^{-1}=\frac{n_{c}\hbar^{2}\pi}{4mp}\int d\theta\sin\theta\left|S^{+}-S^{-}\right|^{2}\cos^{n}\theta.\end{array}\label{eq:taus}\end{alignat}

The second group,\begin{equation}
\begin{array}{c}
\partial_{t}\zeta+\frac{v}{3}\nabla\cdot\boldsymbol{\xi}=-\left\langle l_{z}^{2}\right\rangle \tau_{s0}^{-1}\zeta-\imath\omega\zeta\\
v\nabla\zeta=-\left(\tau_{p+}^{-1}+\left\langle l_{z}^{2}\right\rangle \tau_{s1}^{-1}\right)\boldsymbol{\xi}-\imath\omega\boldsymbol{\xi},\end{array}\label{eq:rho12_set}\end{equation}
 where $\zeta=\rho_{1}+\imath\rho_{2}$ and $\boldsymbol{\xi}=\boldsymbol{\eta}_{1}+\imath\boldsymbol{\eta_{2}}$,
takes into account an average of electron spin precession in the (xy)-plane
with frequency \[
\omega=\left\langle l_{z}\right\rangle \frac{n_{c}\hbar^{2}\pi}{2mp}\int\textrm{Im}\left(S^{+*}S^{-}\right)\sin\theta d\theta+\omega_{B}.\]

Eq.(\ref{eq:rho03_set}) describes the kinetics of charge ($\rho_{0}$)
and spin ($\rho_{3}$) densities in a macroscopic sample. Without
any electric bias this would be diffusion \cite{key-24} described
by equations\begin{gather}
\left\{ \begin{array}{l}
\partial_{t}\rho_{0}=D_{e}\nabla^{2}\rho_{0}-D_{\Delta}\nabla^{2}\rho_{3}\\
\partial_{t}\rho_{3}=D_{s}\nabla^{2}\rho_{3}-D_{\Delta}\nabla^{2}\rho_{0}-\left(1-\left\langle l_{3}^{2}\right\rangle \right)\tau_{s0}^{-1}\rho_{3}\end{array}\right.,\label{eq:rho03_diffusion}\\
\begin{array}{l}
D_{e}=\frac{v^{2}}{3}\frac{\left(\tau_{p+}^{-1}+\tau_{s1}^{-1}\left(1-\left\langle l_{z}^{2}\right\rangle \right)\right)}{\tau_{p+}^{-1}\left(\tau_{p+}^{-1}+\tau_{s1}^{-1}\left(1-\left\langle l_{z}^{2}\right\rangle \right)\right)-\tau_{p-}^{-2}\left\langle l_{z}\right\rangle ^{2}},\\
D_{s}=\frac{v^{2}}{3}\frac{\tau_{p+}^{-1}}{\tau_{p+}^{-1}\left(\tau_{p+}^{-1}+\tau_{s1}^{-1}\left(1-\left\langle l_{z}^{2}\right\rangle \right)\right)-\tau_{p-}^{-2}\left\langle l_{z}\right\rangle ^{2}},\\
D_{\Delta}=\frac{v^{2}}{3}\frac{\left\langle l_{3}\right\rangle \tau_{p-}^{-1}}{\tau_{p+}^{-1}\left(\tau_{p+}^{-1}+\tau_{s1}^{-1}\left(1-\left\langle l_{z}^{2}\right\rangle \right)\right)-\tau_{p-}^{-2}\left\langle l_{z}\right\rangle ^{2}}.\end{array}\nonumber \end{gather}

A homogeneous solution of Eq.(\ref{eq:rho03_set}) in the presence
of electric bias produces the current-field relation for both charge
and spin components $\mathbf{j}_{e}=\sigma_{e}\mathbf{E}$ and $\mathbf{j}_{s}=\sigma_{s}\mathbf{E}$
, which determines electric(\textit{e}) and spin(\textit{s}) conductivities,\begin{equation}
\sigma_{e}=e^{2}\nu\left(\epsilon_{F}\right)D_{e},\,\sigma_{s}=-\left(\hbar/2\right)e\nu\left(\epsilon_{F}\right)D_{\Delta}.\label{eq:sigmas}\end{equation}
A magnetic field dependence of average cluster polarisation $\left\langle l_{z}\right\rangle $
and mean square $\left\langle l_{z}^{2}\right\rangle $ generates
a finite magneto-resistance (MR) of a composite material,\begin{equation}
\Delta\left(B\right)=\frac{\sigma_{e}\left(0\right)}{\sigma_{e}\left(B\right)}-1.\label{eq:delta_def}\end{equation}

The resistance change which can be achieved across an {}``infinite''
magnetic field interval (providing a complete uniaxial polarisation
of all clusters) is given by \begin{equation}
\Delta_{max}=\left(\tau_{p+}/\tau_{p-}\right)^{2}.\label{eq:delta_max}\end{equation}
The maximal MR effect, $\Delta_{max}$ is the largest when both potential
and magnetic scattering potentials are of the same order, and it is
suppressed by additional non-magnetic scattering processes increasing
$\tau_{p+}^{-1}$ \textit{versus} $\tau_{p-}^{-1}$. 

The entire MR curve can be described in terms of magnetic field and
crystalline anisotropy dependent values of $\left\langle l_{z}\right\rangle $
and $\left\langle l_{z}^{2}\right\rangle $, \begin{eqnarray}
\Delta\left(B\right) & =- & \frac{\left\langle l_{z}\right\rangle ^{2}\Delta_{max}}{1+\left(1-\left\langle l_{z}^{2}\right\rangle \right)X},\,\,\textrm{where}\,\, X=\frac{\tau_{p+}}{\tau_{s1}}.\label{eq:delta_a}\end{eqnarray}

When magnetic clusters represent the only source of scattering in
the material, the parameter $X$ in Eq.(\ref{eq:delta_a}) is determined
by the cluster size and material composition. For large clusters with
radius $r_{cl}\gg h/p_{F}$, where $p_{F}$ is the Fermi momentum
in the metal all relaxation rates are proportional to the clusters'
geometrical cross-section, $\tau_{\alpha}^{-1}=\frac{n_{c}}{m_{N}p}\pi\left(pr_{cl}\right)^{2}A_{\alpha}$,
$\left(\alpha=p+,p-,s1\right)$. Table \ref{cap:material_parameters}
illustrates values of these parameters for various ratios between
the exchange $J$ and Fermi energy $E_{F}^{F}$ in the ferromagnet
and the ratios $m_{F}/m_{N}$ between band masses in the normal and
ferromagnetic metals. Here, we use a simplifying model with parabolic
dispersion in both materials with equal densities of conduction band
electrons. The latter constraint determines a spin-independent band
mismatch $U=E_{F}^{N}-E_{F}^{F}$ at the interface which is taken
into account as a scattering potential. The data in Table I indicate
that the depolarisation parameter $X$ is not large ($X\lesssim0.5$).
For small clusters ($pr_{cl}\ll h$), scattering is isotropic, thus
$\tau_{s1}^{-1}$ in the integral in Eq.(\ref{eq:taus}) is suppressed,
as compared to the momentum relaxation rate $\tau_{p+}^{-1}$, thus
making $X\ll1$. Addition of potential scatterers and a contribution
from \mbox{e-phonon} scattering produces a similar effect on the
depolarisation parameter. %
\begin{table}
\begin{tabular}{ccccccc}
\hline 
$J/E_{F}\;$&
$\; m_{F}/m_{N}\;$&
$\quad A_{p+}\quad$&
$\quad A_{p-}\quad$&
$\; A_{s1}\;$&
$\quad\Delta_{\textrm{max}}\quad$&
$\quad X\;$\tabularnewline
\hline 
$0.3$&
$0.5$&
$0.880$&
$0.560$&
$0.450$&
$0.405$&
$0.511$\tabularnewline
$0.3$&
$1.0$&
$0.290$&
$0.040$&
$0.170$&
$0.019$&
$0.586$\tabularnewline
$0.3$&
$2.0$&
$0.510$&
$0.070$&
$0.240$&
$0.137$&
$0.470$\tabularnewline
$0.3$&
$3.0$&
$0.860$&
$0.020$&
$0.370$&
$5.4\cdot10^{-4}$&
$0.430$\tabularnewline
$0.1$&
$0.5$&
$0.740$&
$0.200$&
$0.250$&
$0.073$&
$0.337$\tabularnewline
$0.1$&
$1.0$&
$0.053$&
$0.002$&
$0.030$&
$1.4\cdot10^{-3}$&
$0.566$\tabularnewline
$0.1$&
$2.0$&
$0.480$&
$0.020$&
$0.230$&
$1.7\cdot10^{-3}$&
$0.479$\tabularnewline
$0.1$&
$3.0$&
$0.830$&
$0.010$&
$0.370$&
$1.4\cdot10^{-4}$&
$0.446$\tabularnewline
\hline
\end{tabular}

\caption{\label{cap:material_parameters}Example of estimations of $A_{\alpha}$
in a spherical cluster model for large clusters $pr_{cl}\gg h$ and
strong exchange $J$}
\end{table}

Below, we compare the magnetic field dependence $\Delta\left(B\right)$
for three ensembles of clusters: (a) intrinsically isotropic FmnCs;
(b) identical clusters with colinear easy axes aligned with the external
magnetic field; (c) ensemble of clusters with randomly oriented easy
axes. In all of these three cases we assume the same material composition,
size of clusters, surrounding medium (i.e. fixed values of $\tau_{p\pm}$,
$\tau_{sn}$ and $X$), and thermal equilibrium of angular distribution
of clusters. These three situations will be identified with the following
form of free energy of magnetic subsystems,\begin{equation}
F=-\mu\mathbf{B}\cdot\mathbf{l}+F_{0}=\left\{ \begin{array}{lc}
-\mu Bl_{z} & \;\;\;(a)\\
-\mu Bl_{z}-\alpha l_{z}^{2} & \;\;\;(b)\\
-\mu Bl_{z}-\alpha\left(\mathbf{l}\cdot\mathbf{l_{0}}\right)^{2} & \;\;\;(c)\end{array}\right.\label{eq:free_energies}\end{equation}
where $\mathbf{l_{0}}$ determines the direction of single cluster
easy axis. In cases (a) and (b) the values of $\left\langle l_{z}\right\rangle \equiv\overline{l_{z}}$
and $\left\langle l_{z}^{2}\right\rangle \equiv\overline{l_{z}^{2}}$
are determined by the thermal average\begin{equation}
\overline{A}=N^{-1}\int d\Omega_{\mathbf{l}}Ae^{-\frac{F}{kT}},\: N=\int d\Omega_{\mathbf{l}}e^{-\frac{F}{kT}},\label{eq:average}\end{equation}
where $T$ is temperature. For ensemble (c) with isotropic distribution
of easy axes, an additional averaging over directions of vector $\mathbf{l_{0}}$
in Eq.(\ref{eq:free_energies}c) is needed $\left\langle l_{z}\right\rangle =\left\langle \left\langle \overline{l_{z}}\right\rangle \right\rangle $
and $\left\langle l_{z}^{2}\right\rangle =\left\langle \left\langle \overline{l_{z}^{2}}\right\rangle \right\rangle $,
where $\left\langle \left\langle \left(\dots\right)\right\rangle \right\rangle =\int d\Omega_{\mathbf{l_{0}}}\left(\dots\right)/4\pi$. 

For intrinsically isotropic clusters, ensemble type (a), \begin{eqnarray}
\left\langle l_{z}\right\rangle _{a} & = & \left(x\coth x-1\right)/x,\quad x=B\mu/kT,\nonumber \\
\left\langle l_{z}^{2}\right\rangle _{a} & = & \left(x^{2}-2x\coth x+2\right)/x^{2}.\label{eq:averages_a}\end{eqnarray}
The resulting form of MR is illustrated in Fig.1 by the curve (a).
The influence of intrinsic easy-axis anisotropy of clusters on MR
depends on the temperature regime and the type of cluster ensemble
(b) or (c). At high temperatures $kT>\alpha$, the existence of an
anisotropy axis common for all clusters results in a small correction
to $\left\langle l_{z}\right\rangle $, $\left\langle l_{z}^{2}\right\rangle $
in ensemble (b), \[
\left\langle l_{z}\right\rangle _{b}=\left\langle l_{z}\right\rangle _{a}+\frac{\alpha}{kT}L_{1}^{b}\left(x\right),\,\left\langle l_{z}^{2}\right\rangle _{b}=\left\langle l_{z}^{2}\right\rangle _{a}+\frac{\alpha}{kT}L_{2}^{b}\left(x\right),\]
thus leading to small corrections to $\Delta\left(B\right)$. Here
$L_{1}^{b}\left(x\right)=\left[2x^{2}\left(\coth^{2}x-1\right)+2x\coth x-4\right]/x^{3}$
and $L_{2}^{b}\left(x\right)=\left[4x^{2}\left(2-\coth^{2}x\right)-16x\coth x+20\right]/x^{4}$.
In ensemble (c), which has an isotropic distribution of magnetic moments
at $B=0$, such a correction is even smaller and appear only in the
second order in $\alpha/kT$<1.

\begin{figure}[t]
\includegraphics[%
  scale=0.33]{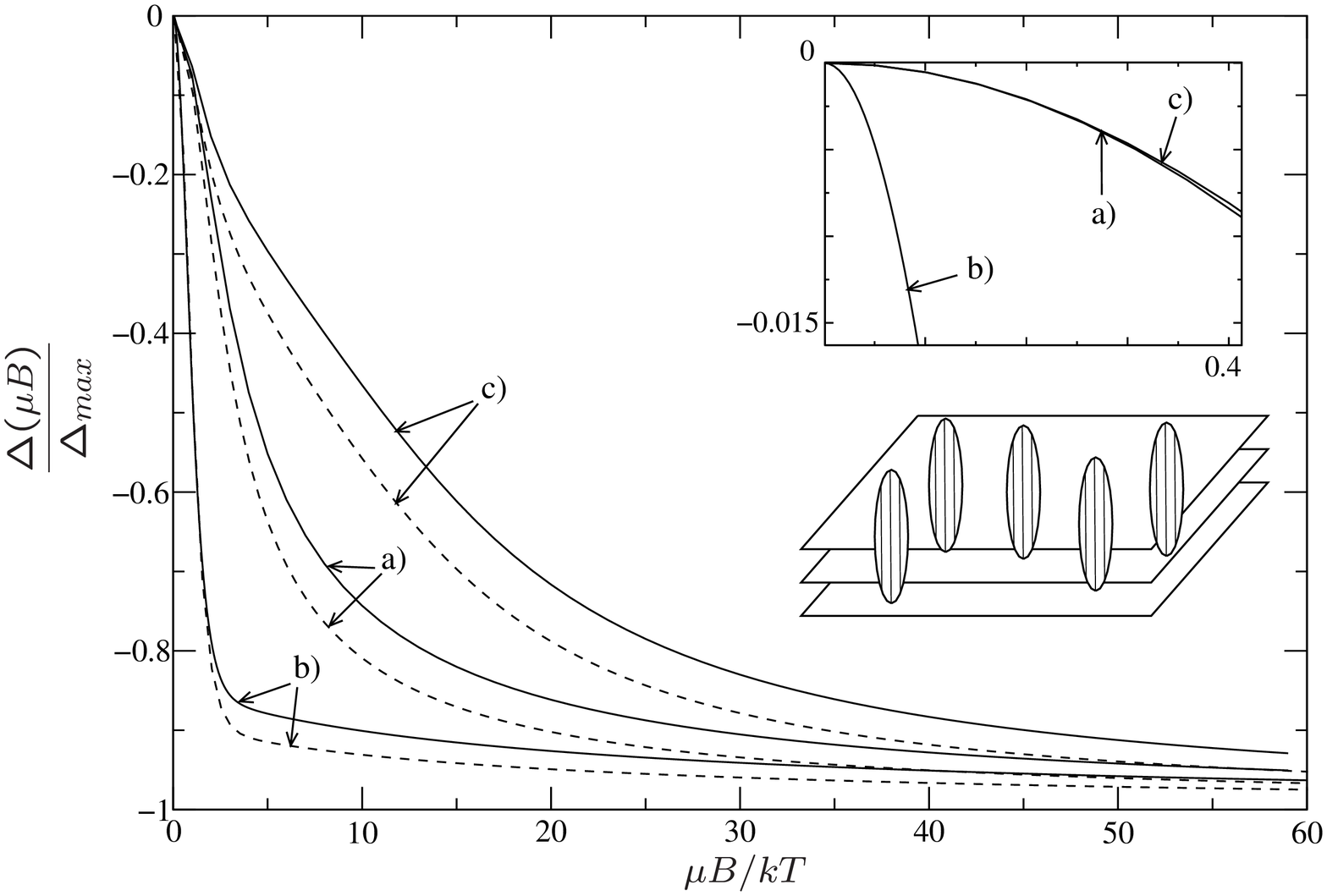}

\caption{\label{cap:high_anisotropy}$\Delta\left(B\right)$ for different
cluster subsystems a) intrinsically isotropic clusters; b) identical
clusters with colinear easy axis in direction of external magnetic
field; c) ensemble of clusters with randomly oriented easy axis; $kT=0.1$,
$\alpha=1$. Solid line shows results for $X=0.5$ and dashed line
for $X=0$.}
\end{figure}
 At low temperatures, $kT<\alpha$, $\Delta\left(B\right)$ was studied
numerically. The obtained results for ensembles (b) and (c) are compared
in Fig.\ref{cap:high_anisotropy} with the MR in ensemble (a). Fig.\ref{cap:high_anisotropy}
shows that a pronounced easy-axis anisotropy enhances the low-field
MR effect, whereas an isotropic distribution of easy axes suppresses
MR via spin relaxation. The inset to Fig.1 demonstrates the low-field
behaviour in all three cases, which complies with the asymptotic expansion
of $\Delta\left(B\right)$ in Eq.(\ref{eq:delta_a}) in powers of
$\mu B/kT\ll1$. Since at small fields $e^{\mu Bl_{z}/kT-F_{0}/kT}\approx e^{-F_{0}/kT}\left(1+\frac{\mu B}{kT}l_{z}\right)$
, \begin{eqnarray*}
\left\langle l_{z}\right\rangle  & \approx & \frac{\mu B}{kT}\frac{\int d\Omega_{\mathbf{l}}l_{z}^{2}\exp\left(-F_{0}/kT\right)}{\int d\Omega_{\mathbf{l}}\exp\left(-F_{0}/kT\right)}=\frac{\mu B}{kT}\left\langle l_{z}^{2}\right\rangle _{B=0}\end{eqnarray*}
 hence the low-field MR is quadratic, \[
\Delta\left(B\right)\approx-\left(\frac{\mu B}{kT}\right)^{2}\frac{\Delta_{max}\left\langle l_{z}^{2}\right\rangle _{B=0}^{2}}{1+\left(1-\left\langle l_{z}^{2}\right\rangle _{B=0}\right)X}.\]
Due to isotropy of FmnC ensembles (a) and (c) at $B=0$, $\left\langle l_{i}l_{j}\right\rangle _{B=0}=\frac{1}{3}\delta_{ij}$
and $\left\langle l_{z}^{2}\right\rangle _{a,B=0}$=$\left\langle l_{z}^{2}\right\rangle _{c,B=0}=\frac{1}{3}$,
whereas \[
\left\langle l_{z}^{2}\right\rangle _{b,B=0}=f\left(\alpha/kT\right),\, f\left(y\right)=\frac{\imath e^{y}y^{-1/2}}{\textrm{erf}\left(\imath y^{-1/2}\right)}-\frac{1}{y}.\]
The low-field MR in each of those three cases is \[
\frac{\Delta\left(B\right)}{\Delta_{max}}=-\left(\frac{\mu B}{kT}\right)^{2}\left\{ \begin{array}{cl}
\left(9+6X\right)^{-1} & \quad(a,c)\\
\frac{f^{2}\left(\alpha/kT\right)}{1+\left[1-f\left(\alpha/kT\right)\right]X} & \quad\;\;(b)\end{array}\right..\]

In summary, we propose the theory of classical magnetoresistance in
a normal metal with embedded ferromagnetic nanoclusters. Our analysis
shows that in a thermodynamic equilibrium the MR effect is largest
when all clusters have the same intrinsic easy-axis anisotropy. Thus
it may be advantageous for applications to engineer anisotropy as
sketched in Fig.\ref{cap:high_anisotropy}, despite the inconvenience
which may be caused by hysteresis in their magnetisation dynamics.

This work was funded by Lancaster-EPSRC Portfolio Parnership and EU
Project {}``SFINX'' NMP2-CT-2003-505587.


\begin{thebibliography}{1}
\bibitem{key-26}G.A. Prinz, Phys. Today \textbf{48}, 58(1995); Science \textbf{282},
1660(1998)
\bibitem{key-11}M.Jamet \textit{et al.}, Phys. Rev. Lett. \textbf{86}, 4676(2001);
F. Parent \textit{et al.}, Phys. Rev. B \textbf{55}, 3683(1997); J.
Meldrin \textit{et al.,} J. Appl. Phys. \textbf{87}, 7013(2000)
\bibitem{key-30}S. Zhang, Appl. Phys. Lett \textbf{61}, 1855(1992); S. Zhang, P. Levy.
J.Appl. Phys. \textbf{73}, 5315(1993)
\bibitem{key-29}A. Berkowitz \textit{et al}., Phys. Rev. Lett. \textbf{68},3745(1992);
J. Xiao \textit{et al.}, Phys. Rev. B \textbf{46}, 9266(1992)
\bibitem{key-34}I. Billas, A. Chatelain, A de Heer, J. Mang. Mang. Mater. \textbf{168},
64(1997); M. Jamet \textit{et al.}, Phys. Rev. B \textbf{61}, 493(2001);
Phys. Rev. B \textbf{69}, 024001(2004)
\bibitem{key-35}K. W. Edmonds \textit{et al.}, Phys. Rev. B \textbf{60}, 472(1999)
\bibitem{key-33}F. Parent \textit{et al.}, Phys. Rev. B \textbf{55}, 3683(1997); P.
Allia \textit{et al.}, Phys. Rev. B \textbf{63}, 180404(R) 
\bibitem{key-24}Eq.(\ref{eq:rho12_set}) is pair of diffusion equations for in plane
kinetic\[
\begin{array}{l}
\partial_{t}\zeta=D_{\perp}\nabla^{2}\zeta-\imath D'_{\perp}\nabla^{2}\zeta-\left(1-\left\langle l_{x}^{2}\right\rangle \right)\tau_{s0}^{-1}\zeta-\imath\omega\zeta,\\
D_{\bot}=\frac{v^{2}}{3}\frac{\left(\tau_{+}^{-1}+\left(1-\left\langle l_{1}^{2}\right\rangle \right)\tau_{s1}^{-1}\right)}{\left(\tau_{+}^{-1}+\left(1-\left\langle l_{1}^{2}\right\rangle \right)\tau_{s1}^{-1}\right)^{2}+\left(\left\langle l_{3}\right\rangle \omega\right)^{2}}\\
D'_{\bot}=\frac{v^{2}}{3}\frac{\left\langle l_{3}\right\rangle \omega}{\left(\tau_{+}^{-1}+\left(1-\left\langle l_{1}^{2}\right\rangle \right)\tau_{s1}^{-1}\right)^{2}+\left(\left\langle l_{3}\right\rangle \omega\right)^{2}}\end{array}\]

\end{thebibliography}
\end{document}